\documentstyle[aps,prb,multicol,psfig]{revtex} 
\newif\ifmynarrow
\mynarrowfalse
\renewcommand{\narrowtext}{%
  \ifmynarrow\hspace*{\fill}\raisebox{-1ex}[0pt][0pt]{%
    \rule{0.3pt}{1ex}%
    \rule[1ex]{20.5pc}{0.3pt}}\fi \mynarrowtrue
  \vspace{-1.0ex}%
  \begin{multicols}{2}%
  \par\global\columnwidth20.5pc
  \global\hsize\columnwidth\global\linewidth\columnwidth
  \global\displaywidth\columnwidth}
\renewcommand{\widetext}{%
  \end{multicols}%
  \vspace{-2.5ex}%
  \noindent\raisebox{1ex}[0pt][0pt]{%
    \rule{20.5pc}{0.3pt}%
    \rule{0.3pt}{1ex}}%
  \par\global\columnwidth42.5pc
  \global\hsize\columnwidth\global\linewidth\columnwidth
  \global\displaywidth\columnwidth}
\begin{document}
\draft 
\tighten

\title{Microwave strengths to induce magnetoresistance oscillations
in high-mobility 2DES\\ in a photon-assisted impurity scattering model}
\author{X.L. Lei}
\address{Department of Physics, Shanghai Jiaotong University,
1954 Huashan Rd., Shanghai 200030, China}

\date{\today}
\maketitle



\vspace*{0.3cm}

\narrowtext
Although the major feature (such as period and phase) of the microwave-induced  
magnetoresistance oscillations in two-dimensional (2D) electron systems is 
insensitive to the behavior of the elastic scattering 
in the photon-assisted impurity scattering model,\cite{Lei-liu} 
the required microwave intensity to induce an effective oscillation depends 
strongly on the form of the impurity potential. 
We present here the calculated results of the longitudinal resistivity $R_{xx}$ 
versus $\omega/\omega_c$ ($\omega_c$ is the cyclotron frequency) of 2D
electrons at lattice temperature $T=1$\,K subjected to microwave irradiations of 
frequency $\omega/2\pi=100$\,GHz in GaAs-AlGaAs heterojunctions having same carrier 
density $N_{\rm e}=3\times 10^{11}$cm$^{-2}$, same zero-magnetic-field linear 
mobility $\mu_0=2.4\times 10^{7}$m$^{2}$V$^{-1}$s$^{-1}$, 
and same level broadening parameter $\alpha=12$, 
but the elastic scatterings being respectively due to: (a) remote charged 
impurities located at a distance $s=60$\,nm from the interface (Fig.\,1); (b) remote 
changed
impurities at $s=20$\,nm from the interface (Fig.\,2); (c) background charged 
impurities (Fig.\,3); (d) short-range ($\delta$-function type) potential (Fig.\,4). 
The microwaves are assumed to be
linearly polarized along the same direction as the dc current ($x$-direction)
having different electric field amplitudes $E_s$ as indicated in the
figures. It is seen that for inducing a roughly equivalent oscillation 
the required microwave strength may be an order of magnitude smaller in the case of 
background impurity scattering than that of remote impurity scattering of 
$s=60$\,nm.
  
\vspace{0.3cm}

\begin{figure}[htb]
\hspace{0.3cm}
  \psfig{figure=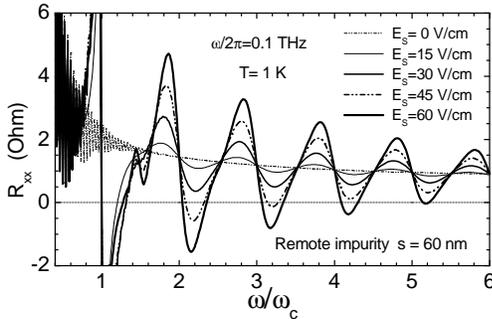,width=6.5cm,height=4.2cm,angle=0,clip=on}
   \caption{The longitudinal magnetoresistivity $R_{xx}$ of a GaAs-based 2DEG 
subjected to a microwave field $E_s\sin(\omega t)$ in the $x$-direction. 
The parameters are: temperature 
$T=1$\,K, electron density $N_{\rm e}=3.0\times 10^{11}$\,cm$^{-2}$, 
zero-magnetic-field linear mobility $\mu_0=2.4\times 10^7$\,cm$^2$ V$^{-1}$ s$^{-1}$, 
and the broadening coefficient $\alpha=12$. The elastic potential is due to remote 
charged impurities located at a distance $s=60$\,nm from the interface.}
\end{figure}
\begin{figure}[htb]
\hspace{0.2cm}
  \psfig{figure=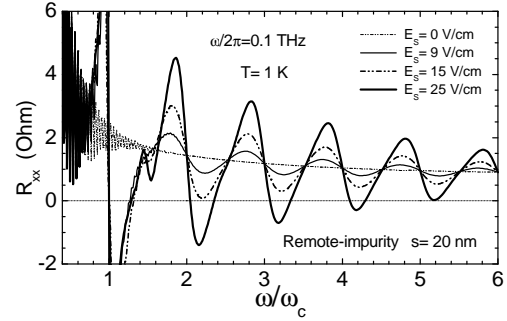,width=6.5cm,height=4.2cm,angle=0,clip=on}
   \caption{Remote imputity scattering, $s=20$\,nm.}
\end{figure}
\begin{figure}[htb]
\hspace{0.3cm}
  \psfig{figure=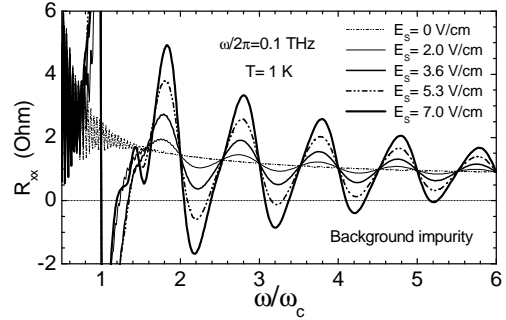,width=6.5cm,height=4.2cm,angle=0,clip=on}
   \caption{Background impurity scattering.}
\end{figure}

\begin{figure}[htb]
\hspace{0.3cm}
  \psfig{figure=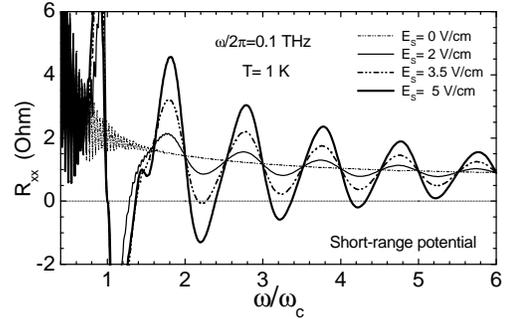,width=6.5cm,height=4.2cm,angle=0,clip=on}
   \caption{Short-range potential scattering.}
\end{figure}

\end{multicols}
\end{document}